\documentclass[aps,twocolumn,pra,tightenlines,floatfix,showpacs,superscriptaddress]{revtex4-1}
\usepackage{graphicx}
\usepackage{amsmath}
\usepackage{amssymb}
\usepackage{times}
\usepackage[english]{babel}

\newcommand{\bq}{\begin{equation}}
\newcommand{\eq}{\end{equation}}
\newcommand{\ba}{\begin{eqnarray}}
\newcommand{\ea}{\end{eqnarray}}

\newcommand{\cD}{{\cal D}}
\newcommand{\calL}{{\cal L}}

\begin{document}

\title{Large-$N$ approximation  for  single- and two-component dilute Bose gases}

\author{Chih-Chun Chien}
\affiliation{Theoretical Division, Los Alamos National Laboratory, MS B213, Los Alamos, NM 87545, USA}
\email{chihchun@lanl.gov}
\author{Fred Cooper}
\affiliation{Theoretical Division, Los Alamos National Laboratory, MS B213, Los Alamos, NM 87545, USA}
\affiliation{Santa Fe Institute, Santa Fe, New Mexico 87501, USA}
\email{cooper@santafe.edu}
\author{Eddy Timmermans}
\affiliation{Theoretical Division, Los Alamos National Laboratory, MS B213, Los Alamos, NM 87545, USA}
\email{eddy@lanl.gov}

\date{\today}

\begin{abstract}
We discuss the mean-field theories obtained from the leading order in a large-$N$ approximation for one- and two- component dilute Bose gases. For a one-component Bose gas this approximation has the following properties: the Bose-Einstein condensation (BEC) phase transition is second order but the critical temperature $T_c$ is not shifted from the non-interacting gas value $T_0$.  The spectrum of excitations in the BEC phase resembles the Bogoliubov dispersion with the usual  coupling constant replaced by the running coupling constant which depends on both temperature and momentum.  We then study  two-component Bose gases with both inter- and intra- species interactions and focus on the stability of the mixture state above $T_c$. Our mean-field approximation predicts  an instability from the mixture state to a phase-separated state when the 
 ratio of the inter-species interaction strength to  the intra-species interaction strength (assuming equal strength for both species) exceeds a critical value. At high temperature this is a structural transition and the global translational symmetry is broken. Our work complements previous studies on the instability of the mixture phase in the presence of BEC.
\end{abstract}


\maketitle
\section{Introduction}
With the capability of tuning inter-particle interactions and
the realization of superfluid phases (see \cite{BlochRMP,ChinRMP} for reviews), cold atoms offer a
window of unparalleled promises onto many-body physics.
While the cold atom prospect of studying quantum criticality has
attracted much attention \cite{SachdevNatPhys}, the finite temperature thermodynamics
\cite{AndersenReview} offers an equally fertile ground for explorations of 
fundamental questions. The power of effective field theory has also proven
to be useful in treating other aspects of many-body physics in cold atoms \cite{Braaten07}. 
Recently developed technologies
for accurate temperature determination \cite{Weld09} and for creating
stable, flat (bulk-like) trapping potentials \cite{ring2} and ring-shape potentials \cite{ring1} provide some
of the necessary tools for probing thermodynamics 
and phase transitions of cold atoms.  Phase separation, the demixing 
phenomenon that spontaneously breaks the global translational symmetry, 
was the second transition after Bose-Einstein condensation (BEC) 
to be observed in the cold-atom laboratory \cite{Stenger98}.  The phase
separation transition that provides a paradigm of second-order
scaling physics in ordinary finite temperature phase transitions 
was observed in a mixture of dilute BECs near zero temperature.
With cold atom technology, the dynamics of a zero temperature 
miscible-immiscible transition of bosonic superfluid mixtures, first 
discussed in studies motivated by the plans of creating liquid 
$^{4}$He-$^{6}$He mixtures \cite{FetterJLTP78}, can now be studied in trapped atoms \cite{EddyPRL98}.  The magnetically controlled Feshbach resonance \cite{ChinRMP,2BECexp1,2BECexp2} 
provides a direct trigger and promises a useful probe of the interaction 
dependence of the phases and phase boundaries. Here
we describe the finite temperature phase separation transition 
above the critical temperature $T_{c}$ of BEC 
in a two-component boson mixture.  

The thermodynamic descriptions of cold atoms encountered
fundamental challenges posed by the inherent limitations of 
the gapless and conserving approximations \cite{Hohenberg65} of interacting bosons.  As
a consequence, the treatments of the BEC-transition in a
single component boson gas generally obtain a transition that
is first order whereas it is known to be second order \cite{AndersenReview}.  The
long-standing problem of the interaction dependence of 
the $T_{c}$ of a single component BEC (see \cite{AndersenReview} for a review) was then discussed 
in the low density limit by intricate reasoning tailored to
the computation of $T_{c}$ (only) \cite{Braaten02,BaymPRL09}.  More recently \cite{LOAFPRL,LOAFlong}, we
developed an auxiliary field description that can overcome the
obstacles of the conserving/gapless approximations and reproduce
the correct order of the BEC-transition at the mean-field level.
This formalism, the Leading Order Auxiliary Field (LOAF) approximation,
introduces two composite fields, one to describe the density $\phi^{*} \phi$
and one to describe the anomalous density $\phi \phi$, where $\phi$ denotes the bosonic field.  This treatment
also predicted a low density limit of the $T_{c}$-dependence
upon the scattering length consistent with previous work based on large-$N$ expansions \cite{LOAFPRL} while providing a complete description  
from which all thermodynamic quantities can be computed \cite{LOAFlong}.  Above
the critical temperature, where the anomalous density vanishes,
this description becomes identical to the usual large-N approximation \cite{MosheLargeN}.
Below, we describe a two-component Bose gas in the large-N approximation and investigate the stability of the 
homogeneous (mixture) phase above the BEC transition temperature to the  
leading order.  To provide context and to gauge the performance of this method, 
we also derive and discuss the large-N predictions for the thermodynamics 
of a single-component finite-temperature Bose gas.

Although the relativistic $O(N)$ model for self-interacting bosons has been the subject of many papers \cite{CJP,Root,CooperAnnPhys,Wilson73,*CJT,*Root75}, the non-relativistic
version which introduces $N$  replicas of the $O(2)$ or equivalently $U(1)$ symmetry of the non-relativistic dilute gas effective theory has not been discussed in great detail in the literature. The main use of the large-N expansion in BEC theories has been to discuss a large-N
approximation near the critical temperature $T_c$  as done by Baym et. al. \cite{BaymPRL09}  and Arnold and Tomasik  \cite{ArnoldTomasik} as well as in the work of Braaten and Radescu \cite{Braaten02}  and has been reviewed by Ref.~\cite{AndersenReview}.  In this paper we will discuss the broken symmetry aspects of this problem following the classic paper of Ref.~\cite{CJP}.  For pedagogical  purposes we will use in this paper a more general method of introducing
auxiliary fields discussed by Refs.~\cite{MosheLargeN,moshe} and briefly discuss the (equivalent)  Hubbard-Stratonovich method that we used in \cite{LOAFPRL,LOAFlong}  and which was also used in \cite{CJP}.  The Hubbard-Stratonovich method relies on one having quartic interactions so it may be  appropriate to introduce the atomic physics community to the more general method of introducing auxiliary fields that can be used  for arbitrary polynomial  (as well as non-polynomial) interactions that preserve the replication symmetry 
\cite{CooperPRD03}.
In this paper we will first derive the large-$N$  expansion for a single component Bose gas.  The large-$N$ expansion when evaluated (as we do here) at $N=1$  is equivalent to choosing $\theta = 0$ in the auxiliary-field approach discussed in detail in Ref.~\cite{LOAFlong} so it does not include the anomalous density explicitly. In Refs.~\cite{LOAFPRL,LOAFlong} we introduced a loop counting parameter $\epsilon$, which is identical to the loop counting parameter $1/N$ above $T_c$ where the anomalous density vanishes.  

What we will show is that for a one-species gas of bosons, the leading order in our large-$N$  approximation leads to a non-perturbative (in coupling constant) mean-field theory  with  reasonable features. As in the more sophisticated LOAF treatment, the leading-order large-$N$ approximation also predicts the correct second order BEC transition. Importantly, we will show that  the large-$N$ theory {\it does} lead to a Bogoliubov-like spectrum for temperatures below $T_c$ because of mixing between the  fluctuations of the boson and the composite field when there is a broken symmetry. However, a shortcoming of this expansion is that it does not predict in the leading order in $1/N$  a  shift  in the critical temperature from that of the free gas, a feature that is shared by the Popov approximation \cite{LOAFlong}. This defect is rectified at the mean-field level by also including an auxiliary field for the anomalous condensate as in the LOAF approximation.  Since the LOAF approximation leads to the same result as the large-$N$ approximation above $T_c$ and we are interested in the stability of the mixture phase of a two-component Bose gas above $T_c$, we will study the simpler large-$N$ approximation here, which ignores the contributions from the anomalous density in the condensate regime in the leading order.  As summarized in Ref.~\cite{AndersenReview} once the $1/N$ corrections (and higher order corrections) to the self energy of the 
boson propagator are included, one does find a shift in $T_c$ so that the $1/N$ expansion at higher orders does include the effects of the anomalous density.  Finally, we note that if comparisons with experimental results in an inhomogeneous trap are needed, one may use the local density approximation \cite{PethickBEC} to include possible inhomogeneity effects.

This paper is organized as the following. Section \ref{sec:single_component} presents our large-$N$ approximation for a single-component interacting Bose gas both below and above the BEC transition temperature. The excitation spectrum in the BEC phase will be analyzed in details. Section \ref{sec:two_component} shows our large-$N$ approximation for two-component Bose gases in the mixture state as well as the phase-separated state. A phase transition between the two states is found in the normal phase and we present a phase diagram from our theory. Section \ref{sec:conclusion} concludes our work.

\section{Large-$N$ theory for a single-component Bose gas}\label{sec:single_component}
The partition function of a single-component Bose gas can be given a many-body theory  path-integral representation \cite{Negele,AndersenReview},
\begin{equation}\label{J.pf.e:Zdef}
   Z[V,\mu,\beta]
   =
   \iint \cD \phi \, \cD \phi^{\ast} \,
   e^{ -S[\phi,\phi^{\ast};V,\mu,\beta] } \>,
\end{equation}
where we are using the Matsubara imaginary time formalism. $\beta=1/(k_B T)$, $\mu$ is the chemical potential, and $V$ is the volume of the system.  The Euclidian action $S[\phi,\phi^{\ast};V,\mu,\beta]$ is given by
\begin{equation}\label{eq:actiondef}
   S[\phi,\phi^{\ast};V,\mu,\beta]
   =
   \int [ d x ] \, \calL [\phi,\phi^{\ast};\mu] \>,
\end{equation}
where we have introduced the notation
\begin{equation}\label{J.pf.e:intdx}
   \int [ d x ]
   =
   \int d^3 x \int_0^\beta d \tau \>.
\end{equation}
For a dilute Bose gas the effective field theory for the problem can be describe by the Euclidian Lagrangian density  \cite{AndersenReview} 
\begin{eqnarray}\label{eq:Ldef}
\mathcal{L}&=&\frac{\hbar}{2}[\phi^*\partial_\tau \phi-\phi\partial_\tau \phi^*]-\frac{1}{2}\left[\phi^*\frac{\hbar^2\nabla^2}{2m}\phi+\phi\frac{\hbar^2\nabla^2}{2m}\phi^*\right]- \nonumber \\
&&\mu\phi^*\phi+\frac{1}{2}\lambda(\phi^*\phi)^2 .
\end{eqnarray}
Here $\lambda$ is the bare coupling constant and we will discuss its renormalization later. This Lagrangian density corresponds to the Hamiltonian $H=\int d^{3}x\left[-\frac{1}{2}\left(\phi^*\frac{\hbar^2\nabla^2}{2m}\phi+\phi\frac{\hbar^2\nabla^2}{2m}\phi^*\right)+\frac{1}{2}\lambda(\phi^*\phi)^2\right]$.
In what follows we set $\hbar=1$ and $k_B=1$. To determine the finite-temperature effective potential for the theory we will be interested in the generating functional for the connected correlation functions, $\ln Z[j]$, where 
\begin{equation} \label{Z}
Z[V,\mu,\beta,j]
   =
   \iint \cD \phi \, \cD \phi^{\ast} \,
   e^{ -S[\phi,\phi^{\ast};V,\mu,\beta]+ \int dx j^\star \phi + \phi^\star j } \>,
\end{equation}

The large-$N$  expansion is a combinatoric trick that reorganizes the Feynman diagrams of the theory in a non-perturbative fashion.  First it sums the loops (bubbles) contributing to the scattering amplitude. Calling this bubble  sum a ``composite-field  propagator", one then re-sums the theory by implementing a loop expansion in terms of the number of composite-field propagator loops  in a diagram.  This way of organizing the Feynman diagrams of the theory can be accomplished formally by introducing $N$ copies of the original theory which is equivalent to introducing a ``color"  index with $N$ components,  or in some cases  extending a theory with a $O(1)$ or  $O(2)$ symmetry  to an $O(N)$ symmetry. For the Lagrangian density \eqref{eq:Ldef} one then introduces (see below)  a composite field $\alpha  = \frac{ \lambda}{N}\sum_{n} \phi_n^* \phi_n$ into the theory by introducing formally a ``$1$" into the generating functional $Z[j]$ shown in Eq.~\eqref{Z}  by using a functional expression for the 
delta function enforcing the definition of $\alpha$. Equivalently the composite field can be introduced using a Hubbard-Stratonovitch 
transformation as shown in \cite{CJP,LOAFPRL,LOAFlong}. As shown below this  converts the quartic self interaction into a trilinear interaction that is quadratic in the original $\phi$ field.  This allows one to perform the path integrals over the original fields $\phi_n$ in the generating functional $Z[j]$ exactly while keeping $\alpha$ fixed.  At this stage one could then determine $\alpha$ by using its
definition and obtain the Weiss self-consistent mean-field theory \cite{Weiss}. The beauty of the path integral approach is that this mean-field theory is the first term in a complete resummation of the theory in terms of loops of higher and higher numbers of the composite-field propagators \cite{CooperAnnPhys}.  Having $N$ copies
of the original theory or extending $O(2)$ to $O(N)$ introduces a small parameter $1/N$ into the theory which allows one to perform the remaining path integration over the composite field $\alpha$,  which arises from inserting the formal expression for the 
delta function into the path integral, using Laplace's method (or the method of steepest descent).  

To make this procedure explicit, we first make $N$ copies of the original field \cite{CJP,Negele} by generalizing $\phi\rightarrow \phi_{1}, \cdots, \phi_{N}$, rescale the coupling constant $\lambda \rightarrow \lambda/N$, and define $\Phi=(\phi_1,\phi_1^{*},\cdots,\phi_{N},\phi_{N}^*)^{T}$ and add external sources $J$  so that the generating functional for the correlation functions is given by 
\begin{equation}
Z[J] = \int \left(\prod_{n=1}^{N}\mathcal{D}\phi_{n}\mathcal{D}\phi_{n}^*\right)  e^{ -S[J,   \phi_n,  \phi_n^*]},
\end{equation}
where the action $S$ is given by
\begin{equation}
S=\int [dx] \left[\frac{1}{2}\Phi^{\dagger}\tilde{G}^{-1}_{0}\Phi+\frac{\lambda}{2N}\left(\sum_{n=1}^{N}\phi_n^* \phi_n\right)^2-J^{\dagger}\Phi \right].
\end{equation}
Here $J=(j_1,j_1^*,j_2,j_2^*,\cdots,j_N,j_N^*)^T$ is the source coupled to $\Phi$, $\bar{G}_{0}^{-1}=diag(h^{(+)},h^{(-)},\cdots,h^{(+)},h^{(-)})$ ($N$ identical copies),  $\tilde{G}_{0}^{-1}=\bar{G}_{0}^{-1}-diag(\mu,\mu,\cdots,\mu,\mu)$ is the bare (noninteracting) Green's function, and $h^{(\pm)}=\pm\partial_{\tau}-\nabla^{2}/(2m)$. The classical value of the $n$-th field is $\phi_{n,c}=(1/Z)(\delta Z/\delta j_n^*)$. Details of the large-N approach and its applications to other fields can be found in Refs.~\cite{MosheLargeN,BickersLargeN,Zeebook}

We then introduce the  auxiliary field  $\alpha = \frac{\lambda}{N} \sum_{n=1}^{N}\phi_n^* \phi_n$ to facilitate our resummation scheme outlined above by  inserting the following  identity inside the path 
integral for the generating functional \eqref{J.pf.e:Zdef} using a formal integral representation of the Dirac delta function  
\begin{eqnarray}\label{eq:delta}
1&=& \int \mathcal{D}\alpha \delta(\alpha-\frac{\lambda}{N}\sum_{n=1}^{N}\phi_n^* \phi_n) \nonumber \\ 
&=& \mathcal{N}\int\mathcal{D}\chi \mathcal{D}\alpha \exp \left[\frac{N}{\lambda}\chi(\alpha-\frac{\lambda}{N}\sum_{n=1}^{N}\phi_n^* \phi_n)  \right] .
\end{eqnarray}
Here $\mathcal{N}=1/(2\pi i)$ is a normalization factor and the $\chi$ integration contour runs parallel to the imaginary axis as discussed in Ref.~\cite{MosheLargeN}. This representation allows one to replace $\sum_{n=1}^{N}\phi_n^*\phi_n$ by $(N/\lambda)\alpha$ in $S$ inside the path integral. Let $G_{0}^{-1}\equiv\bar{G}_{0}^{-1}+diag(\chi,\chi,\cdots,\chi,\chi)$.  It is now possible to perform the quadratic integral over $\phi_n$ exactly to obtain a new
effective action that (because of the large factor $N$) can be evaluated by Laplace's method.
After integrating out $\phi_n$, one has
\bq
Z[J,S,K] = \int\mathcal{D}\chi \mathcal{D}\alpha  e^{-S_{eff}},
\eq
where we have added sources for the auxiliary fields  $\chi$ and $\alpha$ and  
\begin{eqnarray}
S_{eff}&=&\int [dx] \left[-\frac{1}{2}J^{\dagger}G_{0}J-\frac{N}{\lambda}\mu\alpha+\frac{N}{2\lambda}\alpha^2-\frac{N}{\lambda}\chi\alpha  + \right. \nonumber \\
& & \left. \frac{1}{2}Tr\ln G_{0}^{-1}-(S\chi+K\alpha) \right].
\end{eqnarray}
The $Tr\ln G_{0}^{-1}$ term comes from the Gaussian  integration over the bosonic fields  (see Eq.~\eqref{gauss}). Note that the first term and the $Tr\ln G_{0}^{-1}$ term are just $N$ copies of the $U(1)$ theory so they are of order $N$.  We can also rescale the sources $S$ and $K$ to be 
proportional to $N$ so that a large parameter $N$ is in front of the entire action.  This enables us to evaluate the remaining integrals over
$\chi$ and $\alpha$ by Laplace's method (or the stationary phase approximation).  The resulting expansion is a loop expansion in the composite field propagators for $\chi$ and $\alpha$ \cite{CooperAnnPhys}.

The leading order in large-$N$ expansion is obtained by just keeping the contribution to $Z$ evaluated at the minimum of the effective action $S_{eff}$  (i.e. the stationary phase contribution) \cite{CJP,Root}.
 Note that $\delta S_{eff}/\delta j_n^*=-\phi_{n,c}$.  Although we are interested in the theory with $N=1$, for many problems the large-N
 expansion (which is an asymptotic expansion)  gives qualitatively good results at  $N=1$ at leading order and the corrections at next order  bring one closer to the exact answer.  This was seen in the calculation of $T_c$ (in a slightly different context) in Ref.~\cite{ArnoldTomasik}.  

Exactly the same large-N expansion can be obtained from completing the square in a shifted Gaussian integral.  This is the well known 
Hubbard-Stratonovich transformation which is useful when the interactions are only quartic in nature and is based on the identity (here given for 
multi-dimensional integrals) \cite{Negele}
\begin{eqnarray}  \label{gauss}
&& \int  \frac {dx_1 dx_2  \ldots dx_n}{(2 \pi)^{n/2}} \exp \left[-\frac{1}{2}\sum_{i,j} x_i M_{ij} x_j + \sum_{i}x_i j_i\right] =  \nonumber \\
&&[\det M]^{-1/2} \exp\left[\frac{1}{2}\sum_{i,j} j_i M_{ij}^{-1}  j_j \right].
\end{eqnarray}
On a lattice, with the substitutions $j_i \rightarrow  \phi^\star(i) \phi(i )$  $M_{ij}  \rightarrow \delta_{ij}/ \lambda $,  $x_i \rightarrow  \alpha(i)$
we find that the $\phi$ integral becomes quadratic, but we now have to be able to perform the resulting (path) integration over the composite field $\alpha(x)$, which is again done by the stationary-phase approximation.  The resulting large-N expanded effective action is the same as one obtains using the more general method of introducing the composite field $\alpha$ once one eliminates the Lagrange multiplier field  $\chi(x)$  from the 
problem, as will be shown below.

From the Legendre transform of $S_{eff}$ one obtains the generating functional of the one particle irreducible graphs, which is the grand potential $\Gamma[\phi, \chi, \alpha]$ \cite{Zeebook,Negele,IliopoulosRMP}. Explicitly,
 \begin{equation}
 \Gamma=\int [dx]  (J^\dagger\Phi_{c}+S \chi_c + K \alpha_c)+S_{eff}, 
 \end{equation}
 where now $\Phi_{c}, \chi_c, \alpha_c$ stand for  the expectation values of $\Phi, \chi, \alpha$. We define the effective potential for static homogeneous fields $\phi_n, \chi, \alpha$  as $V_{eff}=\Gamma/NV\beta$. Note that the Legendre transformation introduces the expectation values of $\phi_n$ and $\phi_n^*$ in $\Gamma$ and $V_{eff}$ via $\delta\Gamma/\delta\phi_{n,c}^*=j_n$ or, equivalently, $J=G_{0}^{-1}\Phi$ for the expectation values. Now that we have obtained the leading order approximation, we will set $N=1$ so that we are addressing the real dilute gas which has an $U(1)$ symmetry.  At the leading order we find :
\begin{eqnarray}
V_{eff}&=&\frac{1}{2}\Phi^{\dagger} G_{0}^{-1}[\chi]\Phi-\frac{1}{\lambda}\mu\alpha+\frac{1}{2\lambda}\alpha^2-\frac{1}{\lambda}\chi\alpha+ \nonumber \\
& & \frac{1}{2}Tr\ln G_{0}^{-1}[\chi].
\end{eqnarray}
Here we dropped the subscript $c$ for the expectation value, $\Phi=(\phi,\phi^*)^{T}$ for the $N=1$ case, and $G_{0}^{-1}[\chi]$ has been reduced to a $2\times 2$ matrix which depends on $\chi$. The next order in the $1/N$ expansion involves the Gaussian fluctuations in the auxiliary fields $\alpha, \chi$ and will not be included here.

The broken-symmetry condition is determined from the condition that we have found the true minimum of the effective potential:  $\delta V_{eff}/\delta \phi^*=0$, which becomes $\chi\phi=0$. This imposes the following conditions: (1) In the normal phase $\phi=0$ and a finite $\chi$ is allowed and (2) in the broken-symmetry phase, $\phi$ is finite so $\chi=0$. 

After Fourier transforming and  summing over  the Matsubara frequencies, the last term of $V_{eff}$ becomes $\sum_{k}[\omega_k/2+(1/\beta)\ln(1-e^{-\beta\omega_k})]$, where $\omega_k=\epsilon_k+\chi$ and $\epsilon_k=k^{2}/(2m)$. One can eliminate the Lagrange multiplier field  $\chi$ by using the minimum condition $\delta V_{eff}/\delta\alpha=0$. Explicitly,
\begin{equation}\label{eq:1comp_chi}
\chi=-\mu+\alpha.
\end{equation}
Then one obtains
\begin{eqnarray}\label{eq:unrenorm_V}
V_{eff}&=&(-\mu+\alpha)\phi^*\phi-\frac{1}{2\lambda}\alpha^2+ \nonumber \\
& & \sum_{k}[\frac{\omega_k}{2}+\frac{1}{\beta}\ln(1-e^{-\beta\omega_k})].
\end{eqnarray}
Here $\omega_k=\epsilon_k-\mu+\alpha$.

We need to renormalize the theory because Eq.~\eqref{eq:unrenorm_V} is ultra-violet divergent. The renormalized coupling constants can be defined from the effective potential and it is  the value of the scattering amplitude at zero energy- and momentum- transfer or equivalently $ 1/\lambda_R = \delta^2 V_{eff}/\delta\alpha\delta\alpha=1/\lambda+$ (finite polarization terms). The polarization terms can be shown to vanish at zero temperature, so if we define the renormalized coupling constant at $T=0$ then $\lambda_R= \lambda$, and  one may write $\lambda=4\pi\hbar^{2}a/m$, where $a$ is the $s$-wave scattering length at zero temperature.

The renormalization of the chemical potential $\mu$  can be seen more clearly if we rewrite $V_{1,eff}$ in terms of $\chi$ for the moment. The unrenormalized effective potential is given by
\begin{eqnarray}
V_{eff}&=&V_{0}+\chi\phi^*\phi-\frac{1}{2\lambda}(\chi+\mu)^2+ \nonumber \\
& & \sum_{k}[\frac{\epsilon_k+\chi}{2}+\frac{1}{\beta}\ln(1-e^{-\beta\omega_k})].
\end{eqnarray}
Here $V_{0}$ is the unrenormalized  vacuum energy. 
In the classical theory $-\partial V_{eff}/\partial \chi = \mu/\lambda$.  So defining $\mu_R/\lambda$ for the quantum theory  via
\begin{equation}
 -\frac{\partial V_{eff}}{ \partial \chi} = \frac{\mu_R}{\lambda}
 \end{equation} 
 and only keeping the infinite contributions from the quantum fluctuations, we find
\begin{equation} 
\frac{\mu_R}{\lambda}=\frac{\mu}{\lambda}-\sum_{k}\frac{1}{2}.
\end{equation}

Finally there are infinite contributions to the potential that are independent of the field values.  These do not contribute to the equations of motion but can be rendered finite by defining a finite  constant  $V_{R,0} $ via  
\begin{equation}
V_{R,0}- \frac{\mu_{R}^{2}} {2\lambda} =V_{0}- \frac{\mu^2}{ 2\lambda}+ \frac{1}{2} \sum_{k} {\epsilon_k}.
\end{equation}
 With our choice of renormalized parameters we obtain for the renormalized effective potential
\begin{eqnarray}\label{eq:V1effU1}
V_{R,eff}&=&V_{R,0}+\chi\phi^*\phi-\frac{1}{2\lambda}(\chi+\mu_R)^2+ \nonumber \\
& & \sum_{k}\frac{1}{\beta}\ln(1-e^{-\beta\omega_k}).
\end{eqnarray}
It is often convenient to change variables and  write everything in terms of $\alpha_R$. Similar to   Eq.~\eqref{eq:1comp_chi} we introduce $\alpha_R=\chi+\mu_R$. The renormalized effective potential density becomes 
\begin{equation}
V_{eff}=(-\mu+\alpha)\phi^*\phi-\frac{\alpha^2}{2\lambda}+\sum_{k}\frac{1}{\beta}\ln(1-e^{-\beta\omega_k}).
\end{equation}
Here we drop the subscript $R$ and the vacuum energy. $\omega_{k}=\epsilon_{k}-\mu+\alpha$.

In the normal phase, $\phi=0$. The equations of state are derived from $\delta V_{eff}/\delta\alpha=0$ and $-\delta V_{eff}/\delta\mu=\rho$. Explicitly,
\begin{eqnarray}
\frac{\alpha}{\lambda}=\sum_{k}n(\omega_{k}), ~~\rho=\sum_{k}n(\omega_k).
\end{eqnarray}
Here $\omega_k=\epsilon_k-\mu+\alpha$ and $n(x)=[\exp(\beta x)-1]^{-1}$ is the Bose distribution function. We define $k_0=\rho^{1/3}$ and use $k_0^{-1}$ as the unit of length. The BEC transition temperature of a non-interacting Bose gas with density $\rho$ is $T_0=2\pi\hbar^2\rho^{2/3}/[\zeta^{2/3}(3/2)k_B m]$ and we use $k_B T_0$ as our unit of energy.

In the broken-symmetry phase, $\phi$ is finite and the condition $\delta V_{eff}/\delta \phi^*=0$ requires that $\mu=\alpha$. Therefore $\omega_k=\epsilon_k$, which is the same as the dispersion of a non-interacting Bose gas. This implies that the single auxiliary field large- $N$ theory used here does not lead to a shift in the critical temperature $T_c$ from that of a noninteracting gas. One may estimate the shift of the critical temperature by including higher-order terms (see Refs.~\cite{MosheLargeN,AndersenReview} and references therein). A more sophisticated mean-field  theory in the BEC phase, the LOAF theory which contains two auxiliary fields, has been studied in Refs.~\cite{LOAFPRL,LOAFlong} and does lead to a shift in $T_c$ in the leading order. Since the LOAF theory leads to the same result as the large-$N$ approximation above $T_c$ and we are interested in studying the stability of the mixture state of a two-component Bose gas above $T_c$, we will confine ourselves to the simpler leading order in the large-$N$ approximation which ignores the contributions from the anomalous density. However, as we shall see below, the large-$N$ theory {\it does} lead to a Bogoliubov-like spectrum below $T_c$. Thus it is a qualitatively reasonable approximation even below $T_c$.
 We next define $\rho_c=\phi^*\phi$ as the condensate density in the broken-symmetry phase and consider a Bose gas of density $\rho$. Then $\delta V_{eff}/\delta\alpha=0$ and $-\delta V_{eff}/\delta\mu=\rho$ give
\begin{eqnarray}
\frac{\alpha}{\lambda}=\rho_c+\sum_{k}n(\omega_{k}), ~\rho=\rho_c+\sum_{k}n(\omega_k). \label{gap}
\end{eqnarray}
Since $\omega_k=\epsilon_k$, the second equation implies that $\rho_c/\rho$ as a function of $T$ is insensitive to $\lambda$ in this theory.

\begin{figure}
  \includegraphics[width=3.2in,clip] {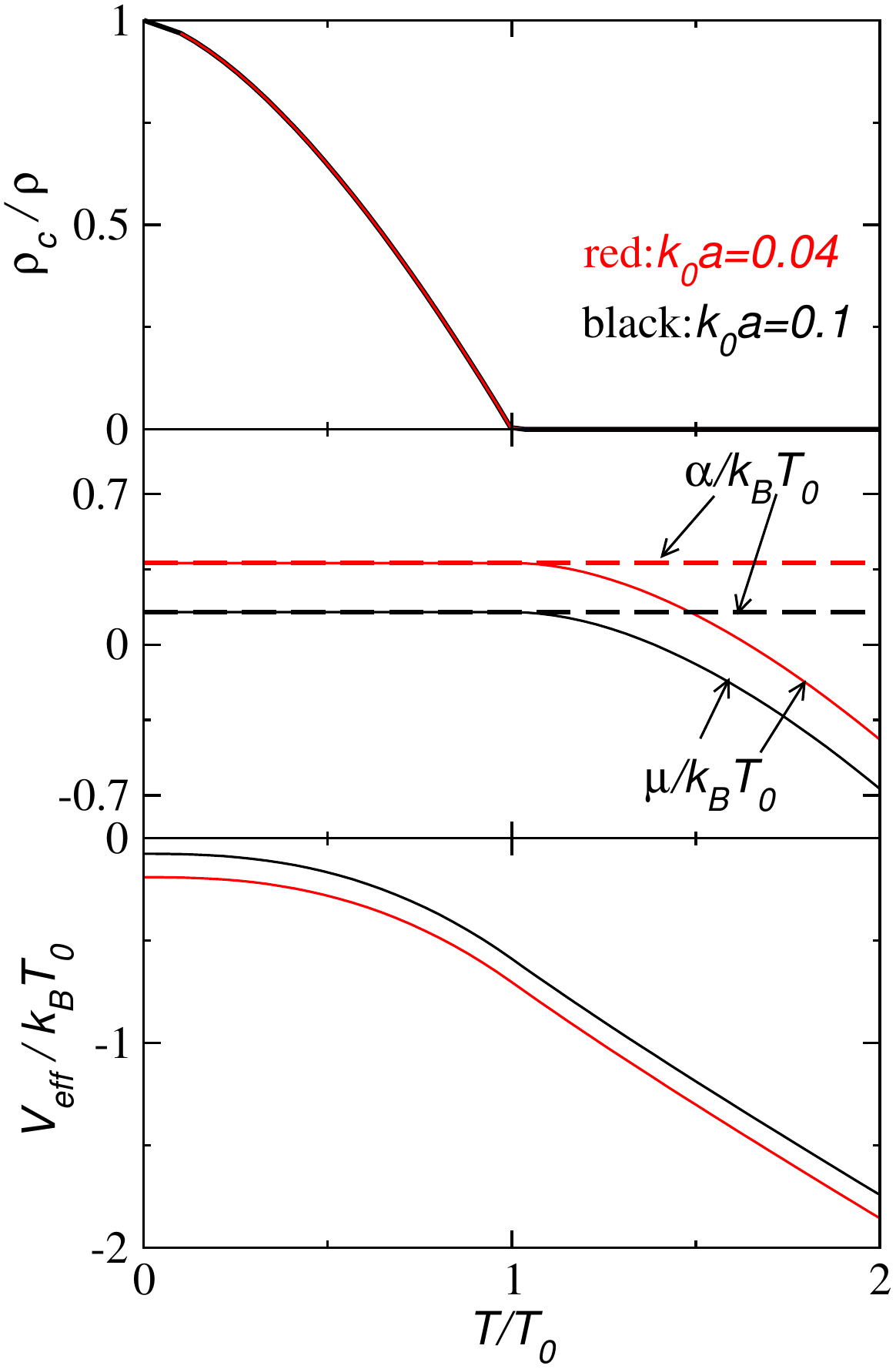}
  \caption{Single auxiliary field large-$N$ theory of a single-component Bose gas. (a) $\rho_c/\rho$ vs. $T$. (b) $\mu$ and $\alpha$ vs. $T$. (c) $V_{eff}$  (evaluated at the minimum) as a function of $T$. Red and black curves correspond to $k_0 a=0.04$ and $0.1$.}
\label{fig:1BEC}
\end{figure}  
Figure~\ref{fig:1BEC} shows $\rho_c$, $\mu$, $\alpha$, and the minimum of $V_{eff}$ as functions of $T$. One important feature is that the BEC transition temperature is fixed at $T_0$ because the dispersion $\omega_k$ is identical to the dispersion of non-interaction bosons regardless of the interaction strength. The transition is second order because $\rho_c$ is continuous at  $T_c=T_0$. The  $V_{eff}$ that is plotted is the value of the effective potential at the minimum where $ \delta V_{eff}/ \delta \phi =\delta V_{eff}/ \delta \alpha =0$. A negative value of $V_{eff}$ corresponds to positive pressure so the system should be mechanically stable.  The BEC condition requires $\alpha=\mu$, which can be verified below $T_c$.

Once we have found the correct ground state of this theory, it is important to calculate the propagators in this ground state.  This has been addressed in detail in the relativistic $O(N)$-model in Ref.~\cite{CJP} and here we will follow a similar procedure in the broken-symmetry phase. In the broken symmetry phase (BEC phase)  the $\phi$ and $\chi$ propagators mix.  To calculate the propagators in the broken symmetry phase one needs to invert the matrix inverse Green's function that is obtained from the effective action which is the generator of all one-particle irreducible graphs and which is the Legendre transform of the generating functional $(-\ln Z)$.  The broken symmetry ground state is described by $\chi = 0$ and $\langle \phi\rangle= \sqrt{\rho_c}>0$, where $\rho_c$ is found from solving Eqs. (\ref{gap})

The  effective action whose static part leads to Eq.~\eqref{eq:V1effU1} is given by
\begin{equation}
\Gamma=\int [dx] \left(\frac{1}{2}\Phi^{\dagger}G^{-1}_{0}\Phi-\frac{N(\chi+\mu)^2}{2\lambda}+ \frac{1}{2}Tr\ln G^{-1}_{0}\right).  \label{action1}
\end{equation}
Here $G^{-1}_{0}[\chi]=diag(-i\omega_n+\omega_k, i\omega_n+\omega_k,\cdots,-i\omega_n+\omega_k, i\omega_n+\omega_k)$ and $\omega_k=\epsilon_k+\chi$.  Since we are interested in the $N=1$ case we will now set $N=1$ and confine ourselves to the actual one-component Bose gas which has a $U(1)$, or equivalently $O(2)$, symmetry.  Let us first look at the $U(1)$ approach.  Here we can use the $U(1)$ symmetry of the theory to choose the vacuum expectation value of $\phi$ to be real. Thus in the broken symmetry phase we
 let $\phi=\sqrt{\rho_c}+\tilde{\phi}$, $\phi^*=\sqrt{\rho_c}+\tilde{\phi}^*$. The term $\chi\phi^*\phi$ becomes $\chi(\rho_c+\sqrt{\rho_c}\tilde{\phi}+\sqrt{\rho_c}\tilde{\phi}^*+\tilde{\phi}\tilde{\phi}^*)$. 
The inverse propagator in the $(\chi,\tilde{\phi})$ sector is not diagonal because of the condensate $\rho_c$. Let $\Psi=(\tilde{\phi},\tilde{\phi}^{*},\chi)^{T}$. Then the fluctuations can be written as $\Psi^{\dagger}\mathcal{D}^{-1}\Psi$. The inverse propagator matrix $\mathcal{D}^{-1}$ is obtained by taking the second derivatives of the effective action and then evaluating these in the broken symmetry ground state where $\chi =0$ and $\langle \phi \rangle = \sqrt{\rho_c}$. 
\begin{eqnarray}
\mathcal{D}^{-1}=\left(\begin{array}{c c c}
\frac{1}{2}\frac{\delta \Gamma_{eff}}{\delta\tilde{\phi}\delta\tilde{\phi}^*} & \frac{\delta \Gamma_{eff}}{\delta\tilde{\phi}\delta\tilde{\phi}}  & \frac{1}{2}\frac{\delta \Gamma_{eff}}{\delta\tilde{\phi}\delta\chi} \\
\frac{\delta \Gamma_{eff}}{\delta\tilde{\phi}^{*}\delta\tilde{\phi}^*} & \frac{1}{2}\frac{\delta \Gamma_{eff}}{\delta\tilde{\phi}^*\delta\tilde{\phi}} & \frac{1}{2}\frac{\delta \Gamma_{eff}}{\delta\tilde{\phi}^*\delta\chi}  \\
\frac{1}{2}\frac{\delta \Gamma_{eff}}{\delta\tilde{\phi}^*\delta\chi} & \frac{1}{2}\frac{\delta \Gamma_{eff}}{\delta\tilde{\phi}\delta\chi} & \frac{\delta \Gamma_{eff}}{\delta\chi\delta\chi}
\end{array}\right).
\end{eqnarray}
The upper $2\times 2$ submatrix is just $(1/2)diag(-i\omega_n+\omega_k,i\omega_n+\omega_k)$, $\frac{\delta \Gamma_{eff}}{\delta\tilde{\phi}\delta\chi}=\sqrt{\rho_c}$, and  
\begin{eqnarray}
\frac{\delta \Gamma_{eff}}{\delta\chi(x)\delta\chi(y)}&=&-\frac{\delta(x-y)}{\lambda}-\frac{1}{2}TrG_{0}\frac{\delta G^{-1}_{0}}{\delta\chi(x)}G_{0}\frac{\delta G^{-1}_{0}}{\delta\chi(y)} \nonumber \\
&=&-\frac{\delta(x-y)}{\lambda}-\frac{1}{2}TrG_{0}(x,y)G_{0}(y,x). \nonumber \\
& &
\end{eqnarray}
Here $x, y$ denote the imaginary time and spatial coordinates, $\delta(x-y)$ is the four-dimensional Dirac delta function, and we have used $\delta G_0/\delta\chi=-\int G_0(\delta G^{-1}_0/\delta\chi)G_0$. After a Fourier transform this becomes
\begin{eqnarray}
\frac{\delta \Gamma_{eff}}{\delta\chi\delta\chi}(K)=-\frac{1}{\lambda}-B(K,T).
\end{eqnarray}
Here $B(K,T)=(1/2)\sum_{Q}TrG_{0}(Q)G_{0}(K+Q)$, $K=(i\omega_n, k)$, $Q=(i\Omega_{\nu},q)$, $\sum_{Q}=(1/\beta)\sum_{\nu}\sum_{q}$ with $\omega_n$ and $\Omega_{\nu}$ being bosonic Matsubara frequencies.  The expression for $B(K,T)$ will be given shortly.

The inverse matrix propagator in the leading order in the broken symmetry ground state is thus
\begin{eqnarray}
\mathcal{D}^{-1}&=&\left(\begin{array}{c c c}
\frac{1}{2}(-i\omega_n+\epsilon_k) & 0  & \frac{1}{2}\sqrt{\rho_c} \\
0 & \frac{1}{2}(i\omega_n+\epsilon_k) & \frac{1}{2}\sqrt{\rho_c}  \\
\frac{1}{2}\sqrt{\rho_c} & \frac{1}{2}\sqrt{\rho_c} & -[\frac{1}{\lambda}+B(K,T)]
\end{array}\right). \nonumber \\
& &
\end{eqnarray}
Here $\chi=0$ in $B(K,T)$. The dispersion relation for $\omega$ is found by setting  $\det\mathcal{D}^{-1}=0$, which yields
\begin{eqnarray}\label{eq:pole}
\left[\frac{1}{\lambda}+B(K,T)\right](\omega_{n}^{2}+\epsilon_k^{2})+\rho_c\epsilon_k=0.
\end{eqnarray}
After the analytical continuation $i\omega_n\rightarrow\omega+i0^{+}$, the solution to $\det\mathcal{D}^{-1}=0$ is
\begin{eqnarray}\label{eq:wbeyond}
\omega^{2}=\epsilon_k\left[\epsilon_k+\lambda(\omega,k,T)\rho_c\right].
\end{eqnarray}
Here $\lambda(\omega,k,T)\equiv \lambda/[1+\lambda B(\omega,k, T)]$ is the running coupling constant.
We will show that $B(\omega, k ,T=0)=0$ so $\lambda(\omega,k,T=0)=\lambda$. Therefore at $T=0$, 
\begin{equation}
\omega^{2}=\epsilon_k\left[\epsilon_k+\lambda\rho_c\right].
\end{equation}
One may compare this with the Bogoliubov dispersion $\omega_{B}^{2}=\epsilon_{k}(\epsilon_k+2\lambda\rho_c)$ and see that the dispersions are similar. The factor of two comes from the fact that we have ignored the contribution from the anomalous density  
$\langle \phi \phi \rangle$ in the lowest order of our large-$N$ approximation. This contribution should get restored at higher order in the expansion. 

These results for the inverse propagator can also be discussed in the $O(2)$ language
by writing $\phi=\phi_1+i\phi_2$ and $\phi^{*}=\phi_1-i\phi_2$.  Here we use the $O(2)$ symmetry to choose the condensate in the $``1"$
direction.  Then,
in the broken symmetry phase $\phi^{(1)}_1=\sqrt{\rho_c}+\sigma$ and $\phi^{(1)}_2=\pi$. 
The inverse  propagator in the $(\chi,\sigma,\pi)$ representation takes now a slightly different form. The condensate density in this case only couples to $\sigma$ but not $\pi$. We define $\bar{\Psi}=(\sigma,\pi,\chi)^{T}$ and the fluctuations are $\bar{\Psi}^{\dagger}\bar{\mathcal{D}}^{-1}\bar{\Psi}$. Using Eq.~\eqref{action1} we obtain 
\begin{equation}
\bar{\mathcal{D}}^{-1}=\left(\begin{array}{c c c}
\epsilon_k & -\omega_n  & \sqrt{\rho_c} \\
\omega_n & \epsilon_k & 0  \\
\sqrt{\rho_c} & 0 & -[\frac{1}{\lambda}+B(K,T)]
\end{array}\right). 
\end{equation}
Note that the time-derivative terms in Eq.~\eqref{action1} becomes $\pi\partial_\tau \sigma-\sigma\partial_\tau\pi$ and this results in off-diagonal elements in the sub-matrix corresponding to $\pi$ and $\sigma$. From $\det\bar{\mathcal{D}}^{-1}=0$ one finds exactly the same dispersion as the one given by Eq.~\eqref{eq:wbeyond}. Therefore the Bogoliubov-like dispersion at $T=0$ emerges when one calculates the propagators in the correct broken symmetry ground state. 

Now we show $B(K,T)$ explicitly. We define $G_{11}=1/(-i\Omega_{\nu}+\omega_q)$ and $G_{22}=1/(i\Omega_{\nu}+\omega_q)$, where $\omega_{q}=\epsilon_q+\chi$. Then $B(K)=(1/2)\sum_{Q}[G_{11}(Q)G_{11}(Q+K)+G_{22}(Q)G_{22}(Q+K)]$. After summing over the Matsubara frequency, $B(K,T)$ becomes  
\begin{equation}
\frac{1}{2}\sum_{q}\left[\frac{n(\omega_{q+k})-n(\omega_q)}{i\omega_n+\omega_{q}-\omega_{q+k}}+\frac{n(\omega_{q})-n(\omega_{q+k})}{i\omega_n-\omega_{q}+\omega_{q+k}}  \right].
\end{equation}
Here we have used $n(\omega_q+i\omega_n)=n(\omega_q)$ and $n(x)+n(-x)=-1$.
Defining $\Delta\omega\equiv\omega_{q+k}-\omega_{k}$ and $\Delta n\equiv n(\omega_{q+k})-n(\omega_q)$, we obtain the result 
\begin{eqnarray}
B(K,T)&=&\frac{1}{2}\sum_{q}\Delta n \left[\frac{1}{i\omega_n-\Delta\omega}-\frac{1}{i\omega_n+\Delta\omega}\right].
\end{eqnarray}
The function $B(\omega, k,T)$ is then evaluated by the analytic continuation $i\omega_n\rightarrow \omega+i0^+$ \cite{Fetterbook}.
One important consequence follows immediately. In the broken-symmetry phase $\omega_{q}=\epsilon_q$ and $n(\omega_q)=0$ as $T\rightarrow 0$. Therefore $B(\omega, k,T=0)=0$ and this leads to a Bogliubov-like dispersion for the gapless mode inferred from the pole of the inverse propagator $\mathcal{D}^{-1}$.

At finite $T$ one can use the identity $1/(x+i0^+)=P(1/x)-i\pi\delta(x)$, where $P$ denotes the Cauchy principle integral, to obtain the full expression:
\begin{eqnarray}
B(\omega, k,T)&=&\frac{1}{2}P\sum_{q}\Delta n\left[\frac{1}{\omega-\Delta\omega}-\frac{1}{\omega+\Delta\omega}\right]+ \nonumber \\
& &i\frac{\pi}{2}\sum_{q}\Delta n[\delta(\omega+\Delta\omega)-\delta(\omega-\Delta\omega)].
\end{eqnarray}
Thus at finite $T$ one has to solve Eq.~\eqref{eq:pole} with $B(\omega, k, T)$ to find the (complex) dispersion.

\section{The normal phase of a two-component Bose gas}\label{sec:two_component}
The effective action of a two-component Bose gas is a generalization of Eqs.~\eqref{eq:actiondef} and \eqref{eq:Ldef}
\begin{eqnarray}
S[\phi_j,\phi^{*}_j]&=&\int [dx]\left\{\sum_{j=1,2}\left[ \frac{\hbar}{2}[\phi^{*}_{j}\partial_{\tau}\phi_{j}-\phi_{j}(\partial_{\tau}\phi^{*}_{j})] -\right.\right. \nonumber \\
& &\frac{1}{2}[\phi^{*}_{j}(\frac{\nabla^{2}}{2m_j}\phi_{j})+\phi_{j}(\frac{\nabla^{2}}{2m_{j}}\phi^{*}_{j})]-\mu_{j}\phi^{*}_{j}\phi_{j}+ \nonumber \\
& &\left.\left. \frac{1}{2}\lambda_{j}(\phi^{*}_{j}\phi_{j})^{2}\right]+\lambda_{12}(\phi^{*}_{1}\phi_{1})(\phi^{*}_{2}\phi_{2})\right\}.
\end{eqnarray} 
We again introduce the large parameter $N$ into the theory by the replication trick $\phi_j \rightarrow \phi_{j,n}$ where $n =1, 2, \ldots N$ and rescale the coupling constants $\lambda_j \rightarrow \lambda_j/N$ and $\lambda_{12}\rightarrow \lambda_{12}/N$. In the following we use a similar set of symbols to denote physical quantities of a two-component Bose gas. This set of symbols should not be confused with those for a single-component Bose gas in the previous discussion.

The action with the source term after the replication becomes
\begin{eqnarray}
S&=&\int [dx] \left[\frac{1}{2}\Phi^{\dagger}\tilde{G}^{-1}_{0}\Phi+\frac{\lambda_1}{2N}\left(\sum_{n=1}^{N}\phi_{1,n}^*\phi_{1,n}\right)^2+ \right. \nonumber \\
& &\frac{\lambda_2}{2N}\left(\sum_{n=1}^{N}\phi_{2,n}^*\phi_{2,n}\right)^2+ \frac{\lambda_{12}}{N}\left(\sum_{n=1}^{N}\phi_{1,n}^* \phi_{1,n}\right)\times  \nonumber \\
& &\left. \left(\sum_{n=1}^{N}\phi_{2,n}^* \phi_{2,n}\right)-J^{\dagger}\Phi \right].
\end{eqnarray}
Here we define $\Phi=(\phi_{1,1},\phi_{1,1}^{*},\phi_{2,1},\phi_{2,1}^{*}, \cdots)^{T}$, $\bar{G}_{0}^{-1}=diag(h_1^{(+)},h_1^{(-)},h_2^{(+)},h_2^{(-)},\cdots)$, $\tilde{G}^{-1}_{0}=\bar{G}_{0}^{-1}-diag(\mu_1,\mu_1,\mu_2,\mu_2,\cdots)$ is the bare (noninteracting) Green's function of a two-component Bose gas, $h_{j}^{(\pm)}=\pm\partial_{\tau}-\nabla^{2}/(2m_j)$ for $j=1,2$. 
There are $N$ copies in $\Phi$, $\bar{G}_{0}^{-1}$, and $\tilde{G}_{0}^{-1}$. $J$ is the source coupled to $\Phi$. The identity 
\begin{eqnarray}
1&=&\mathcal{N}^{2}\int\mathcal{D}\chi_1\mathcal{D}\chi_2\mathcal{D}\alpha_1\mathcal{D}\alpha_2 \exp\left[\frac{N}{\lambda_1}\chi_1(\alpha_1-  \right. \nonumber \\
& &\left. \frac{\lambda_1}{N}\sum_{n=1}^{N}\phi_{1,n}^* \phi_{1,n})+\frac{N}{\lambda_2}\chi_2(\alpha_2-\frac{\lambda_2}{N}\sum_{n=1}^{N}\phi_{2,n}^* \phi_{2,n}) \right] \nonumber \\
& &
\end{eqnarray}
has the effect of introducing two delta functions similar to the case of a single-component Bose gas so one can replace $\sum_{n=1}^{N}\phi_{j,n}^* \phi_{j,n}$ by $(N/\lambda_j)\alpha_j$ in $S$. This replacement facilitates our resummation scheme and we will treat $1/N$ as a small parameter. Let $G_{0}^{-1}\equiv\bar{G}_{0}^{-1}+diag(\chi_1,\chi_1,\chi_2,\chi_2,\cdots)$. After integrating out $\phi_{j,n}$, one has
\begin{eqnarray}
S_{eff}&=&\int [dx] \left[-\frac{1}{2}J^{\dagger}G_{0}J-\frac{N}{\lambda_1}\mu_1\alpha_1-\frac{N}{\lambda_2}\mu_2\alpha_2+\frac{N}{2\lambda_1}\alpha_1^2 +  \right. \nonumber \\
& & \frac{N}{2\lambda_2}\alpha_2^2+\frac{N\lambda_{12}}{\lambda_1\lambda_2}\alpha_1\alpha_2-\frac{N}{\lambda_1}\chi_1\alpha_1-\frac{N}{\lambda_2}\chi_2\alpha_2  + \nonumber \\
& & \left. \frac{1}{2}Tr\ln G_{0}^{-1} -K^{\dagger}X\right].
\end{eqnarray}
Here $X=(\chi_1,\chi_2,\alpha_1,\alpha_2)^{T}$ with its source term $K$  and expectation value $X_c$
As in the single-component case we evaluate the path integrals over $\chi_j, \alpha_j$ via the method of stationary phase or steepest descent
and in the leading order in large-$N$ we keep only the contributions at the stationary phase point.

The generator of the one-particle irreducible diagrams is obtained from the Legendre transform of $S_{eff}$. Explicitly, $\Gamma=\int (J^{\dagger}\Phi_c+ K^{\dagger} X_c+S_{eff})$, where $\Phi_c$ is the classical value of $\Phi$. We define the effective potential as $V_{eff}=\Gamma/NV\beta$.  Keeping the leading term in the $1/N$ expansion, and then setting $N=1$ we obtain the effective potential for static homogeneous fields 
\begin{eqnarray}
V_{eff}&=&\frac{1}{2}\Phi^{\dagger}G_0^{-1}\Phi-\frac{1}{\lambda_1}\mu_1\alpha_1-\frac{1}{\lambda_2}\mu_2\alpha_2+\frac{1}{2\lambda_1}\alpha_1^2+ \nonumber \\
& & \frac{1}{2\lambda_2}\alpha_2^2+\frac{\lambda_{12}}{\lambda_1\lambda_2}\alpha_1\alpha_2-\frac{1}{\lambda_1}\chi_1\alpha_1-\frac{1}{\lambda_2}\chi_2\alpha_2  + \nonumber \\
& & \frac{1}{2}Tr\ln G_{0}^{-1}.
\end{eqnarray}
Here $\Phi=(\phi_1,\phi_1^{*},\phi_2,\phi_2^{*})^{T}$ and $G_{0}^{-1}$ has been reduced to a $4\times 4$ matrix. Again the Legendre transformation introduces the expectation values of $\phi_{j,n}$ and $\phi_{j,n}^*$ to $\Gamma$ and $V_{eff}$ via $J=G_{0}^{-1}\Phi$ for the expectation values. 
 The broken-symmetry condition is determined from the condition that we have found the true minimum of the effective potential:  $\delta V_{eff}/\delta \phi_j^*=0$, which becomes $\chi_j\phi_j=0$. In the normal phase $\phi_j=0$ while in the broken-symmetry phase $\chi_j=0$.  In the normal phase, the first term in $V_{eff}$  is zero at the minimum of the potential (which occurs at $\phi_j =0$). 

In the following we will focus on the normal phase of the mixture state and consider $\rho_1=\rho_2=\rho_0$ and $m_1=m_2=m$, where $\rho_0$ is the density of a non-interacting single-component Bose gas with the BEC transition temperature $T_0=2\pi\hbar^{2}\rho_0^{2/3}/[\zeta^{2/3}(3/2)k_B m]$. Similar to the case of a single-component Bose gas, we define $k_0=\rho_0^{1/3}$ and use $k_0^{-1}$ and $k_B T_0$ as the units of length and energy.

The last term in $V_{eff}$ can be evaluated using the standard Matsubara frequency summation technique and it becomes $\sum_{k,j}[\omega_j/2+(1/\beta)\ln(1-e^{-\beta\omega_j})]$, where $\omega_j=\epsilon_j+\chi_j$ and $\epsilon_j=\hbar^{2}k^{2}/(2m_j)$.
To express $V_{eff}$ as a functional of $\alpha_j$ and $\mu_j$, we use $\delta V_{eff}/\delta \alpha_j=0$ to obtain
\begin{eqnarray}\label{eq:2BEC_chi}
\chi_j=-\mu_j+\alpha_j+\frac{\lambda_{12}}{\lambda_{\bar{j}}}\alpha_{\bar{j}}.
\end{eqnarray}
Here $\bar{j}=1$ if $j=2$ and $\bar{j}=2$ if $j=1$. This leads to 
\begin{eqnarray}
V_{eff}&=&\sum_{j}(-\mu_j+\alpha_j+\frac{\lambda_{12}}{\lambda_{\bar{j}}}\alpha_{\bar{j}})\phi_j^*\phi_j-\frac{1}{2\lambda_1}\alpha_1^2- \frac{1}{2\lambda_2}\alpha_2^2 \nonumber \\
& &-\frac{\lambda_{12}}{\lambda_1\lambda_2}\alpha_1\alpha_2  + \sum_{k,j}\left[\frac{\omega_j}{2}+\frac{1}{\beta}\ln(1-e^{-\beta\omega_j}) \right].
\end{eqnarray}

The renormalization of $V_{eff}$ is similar to the procedure of a single-component Bose gas. Firstly one can show that $\delta^2 V_{eff}/\delta \alpha_{j}\delta \alpha_{j}=-1/\lambda_{j}+$ (finite terms) and $\delta^2 V_{eff}/\delta \alpha_{1}\delta \alpha_{2}=-\lambda_{12}/\lambda_{1}\lambda_{2}+$ (finite terms). This implies that the physical coupling constants only get finite renormalization, and as in the single field case  $\lambda_j$ and $\lambda_{12}$ are equal to their renormalized values at $T=0$. Then one may let $\lambda_{j}=4\pi\hbar^{2} a_{j}/m_j$ and $\lambda_{12}=2\pi\hbar^{2} a_{12}/m_{r}$, where $a_{1}, a_{2}, a_{12}$ are the $s$-wave scattering lengths of the intra- and inter-species collisions at $T=0$  and $m_r=m_1m_2/(m_1+m_2)$ is the reduced mass. To render the theory finite, we only need to consider the (infinite) renormalization of the chemical potential and vacuum energy.

To make this procedure more transparent, we use Eq.~\eqref{eq:2BEC_chi} to express $V_{eff}$ in terms of $\chi_{j}$ for the moment. This gives
\begin{eqnarray}
V_{eff}&=&V_0+\sum_{j}\chi_j\phi_j^*\phi_j+\frac{\lambda_2}{2\bar{\lambda}^{2}}(\chi_1+\mu_1)^2+ \nonumber \\
& &\frac{\lambda_1}{2\bar{\lambda}^{2}}(\chi_2+\mu_2)^2 -\frac{\lambda_{12}}{\bar{\lambda}^2}(\chi_1+\mu_1)(\chi_2+\mu_2)+ \nonumber \\
& & \sum_{k,j}\left[\frac{\epsilon_j+\chi_j}{2}+\frac{1}{\beta}\ln(1-e^{-\beta\omega_j}) \right].
\end{eqnarray}
Here $\bar{\lambda}^{2}\equiv \lambda_{12}^{2}-\lambda_1\lambda_2$ and $\omega_j=\epsilon_j+\chi_j$. The renormalization of $\mu_j$ follows the set of equations
\begin{eqnarray}
\frac{\lambda_2}{\bar{\lambda}^{2}}\mu_1-\frac{\lambda_{12}}{\bar{\lambda}^{2}}\mu_2+\sum_{k}\frac{1}{2}&=&\frac{\lambda_2}{\bar{\lambda}^{2}}\mu_{1R}-\frac{\lambda_{12}}{\bar{\lambda}^{2}}\mu_{2R}, \nonumber \\
\frac{\lambda_1}{\bar{\lambda}^{2}}\mu_2-\frac{\lambda_{12}}{\bar{\lambda}^{2}}\mu_1+\sum_{k}\frac{1}{2}&=&\frac{\lambda_1}{\bar{\lambda}^{2}}\mu_{2R}-\frac{\lambda_{12}}{\bar{\lambda}^{2}}\mu_{1R}.
\end{eqnarray}
This renormalization absorbs the divergent term $\sum_{k,j}(\chi_j/2)$ in $V_{eff}$. Then the vacuum energy is renormalized by
\begin{eqnarray}
& &V_{0}+\frac{\lambda_2}{\bar{\lambda}^{2}}\mu_{1}^{2}+\frac{\lambda_1}{\bar{\lambda}^{2}}\mu_{2}^{2}-\frac{\lambda_{12}}{\bar{\lambda}^{2}}\mu_{1}\mu_{2}+\sum_{k,j}\frac{\epsilon_j}{2}= \nonumber \\
& & V_{0R}+\frac{\lambda_2}{\bar{\lambda}^{2}}\mu_{1R}^{2}+\frac{\lambda_1}{\bar{\lambda}^{2}}\mu_{2R}^{2}-\frac{\lambda_{12}}{\bar{\lambda}^{2}}\mu_{1R}\mu_{2R}.
\end{eqnarray}
This absorbs the divergent term $\sum_{k,j}(\epsilon_j/2)$ so there is no divergence in $V_{eff}$ after the renormalization.

Following Eq.~\eqref{eq:2BEC_chi} we let $\chi_j=-\mu_{jR}+\alpha_{jR}+(\lambda_{12}/\lambda_{\bar{j}})\alpha_{\bar{j}R}$ and rewrite $V_{eff}$ in terms of $\alpha_{jR}$. The renormalized $V_{eff}$ is
\begin{eqnarray}\label{eq:2BEC_Veff}
V_{eff}&=&\sum_{j}(-\mu_j+\alpha_j+\frac{\lambda_{12}}{\lambda_{\bar{j}}}\alpha_{\bar{j}})\phi_j^*\phi_j-\frac{1}{2\lambda_1}\alpha_1^2- \frac{1}{2\lambda_2}\alpha_2^2 \nonumber \\
& &-\frac{\lambda_{12}}{\lambda_1\lambda_2}\alpha_1\alpha_2  + \sum_{k,j}\frac{1}{\beta}\ln(1-e^{-\beta\omega_j}).
\end{eqnarray}
Here we drop the subscript $R$ and the vacuum energy.
We first consider the normal phase of the mixture state. From $\delta V_{eff}/\delta\alpha_j=0$ and $\rho_j=-\delta V_{eff}/\delta \mu_{j}$ we obtain
\begin{eqnarray}\label{eq:2BEC_EOS_mix}
& &\frac{1}{\lambda_j}\alpha_j+\frac{\lambda_{12}}{\lambda_1\lambda_2}\alpha_{\bar{j}}=\sum_{k}\left[n(\omega_j)+\frac{\lambda_{12}}{\lambda_{j}}n(\omega_{\bar{j}}) \right], \nonumber \\
& &\rho_j=\sum_{k}n(\omega_j).
\end{eqnarray}
The solution along with $\phi_j=0$ then determines the extremum of $V_{eff}$. To determine the stability of the mixture state, we compare the results with those obtained from the phase separated state. Since our formalism uses the grand-canonical ensemble, one has to compare different states with the same chemical potential $\mu_j$. The state with lower $V_{eff}$ should be energetically stable. When the two curves of $V_{eff}$ intersect, it signals a phase transition into a different state.

The broken-symmetry phase emerges when $\chi_j$ vanishes according to the condition $\chi_j\phi_j=0$.  By analyzing \eqref{eq:2BEC_EOS_mix} with $\chi_j=-\mu_{j}+\alpha_{j}+(\lambda_{12}/\lambda_{\bar{j}})\alpha_{\bar{j}}$ one can see that this condition determines the critical temperature $T^{mix}_{c,j}$ and for each component it coincides with the BEC transition temperature of a non-interacting Bose gas with the same density. For the case $\rho_1=\rho_2=\rho_0$ and $m_1=m_2=m$,  $T^{mix}_{c,1}=T^{mix}_{c,2}=T_0$, which is independent of $\lambda_{1},\lambda_{2},\lambda_{12}$. One has to include higher order corrections in the large-$N$ theory to get corrections to the transition temperature.

The effective potential and equations of state for the phase-separated state are similar to those of the single-component Bose gas. For one of the species occupying part of the space, its effective potential is 
\begin{eqnarray}
V^{ps}_{eff}&=&(-\mu_j+\alpha^{ps}_{j})(\phi^{ps}_j)^*\phi^{ps}_{j}-\frac{1}{2\lambda_j}(\alpha^{ps}_{j})^2+ \nonumber \\
& & \sum_{k}\frac{1}{\beta}\ln(1-e^{-\beta\omega^{ps}_j}).
\end{eqnarray}
Here $\mu_j$ needs to match the chemical potential of species $j$ in the mixture phase. As a consequence, the density of the phase-separated state will be different from $\rho_j$ so we denote it by $\rho_j^{ps}$. The energy dispersion is $\omega^{ps}_j=\epsilon_j-\mu_j+\alpha^{ps}_{j}$. Since the BEC transition temperature scales as $(\rho^{ps}_{j})^{2/3}$, it is possible that in order to match $\mu_j$, the phase-separated state may enter the broken-symmetry phase. Therefore we show the equations of state of the phase-separated state in the normal phase as well as in the broken-symmetry phase.

In the normal phase, $\phi_j=0$ and 
\begin{eqnarray}
& &\frac{\alpha^{ps}_j}{\lambda_j}=\sum_{k}n(\omega^{ps}_j), ~\rho^{ps}_{j}=\sum_{k}n(\omega^{ps}_j).
\end{eqnarray}
Here $\omega^{ps}_j=\epsilon_j-\mu_j+\alpha^{ps}_{j}$. In the broken-symmetry phase, $\rho^{ps}_{c,j}\equiv (\phi^{ps}_{j})^*\phi^{ps}_{j}$ and one has
\begin{eqnarray}
& &\frac{\alpha^{ps}_j}{\lambda_j}=\rho^{ps}_{c,j}+\sum_{k}n(\omega^{ps}_j), \nonumber \\
& &\rho^{ps}_{j}=\rho^{ps}_{c,j}+\sum_{k}n(\omega^{ps}_j).
\end{eqnarray}
The dispersion is $\omega^{ps}_{j}=\epsilon_j$ due to the broken-symmetry condition $\chi_j=-\mu_j+\alpha_j=0$.

\begin{figure}
  \includegraphics[width=3.in,clip] {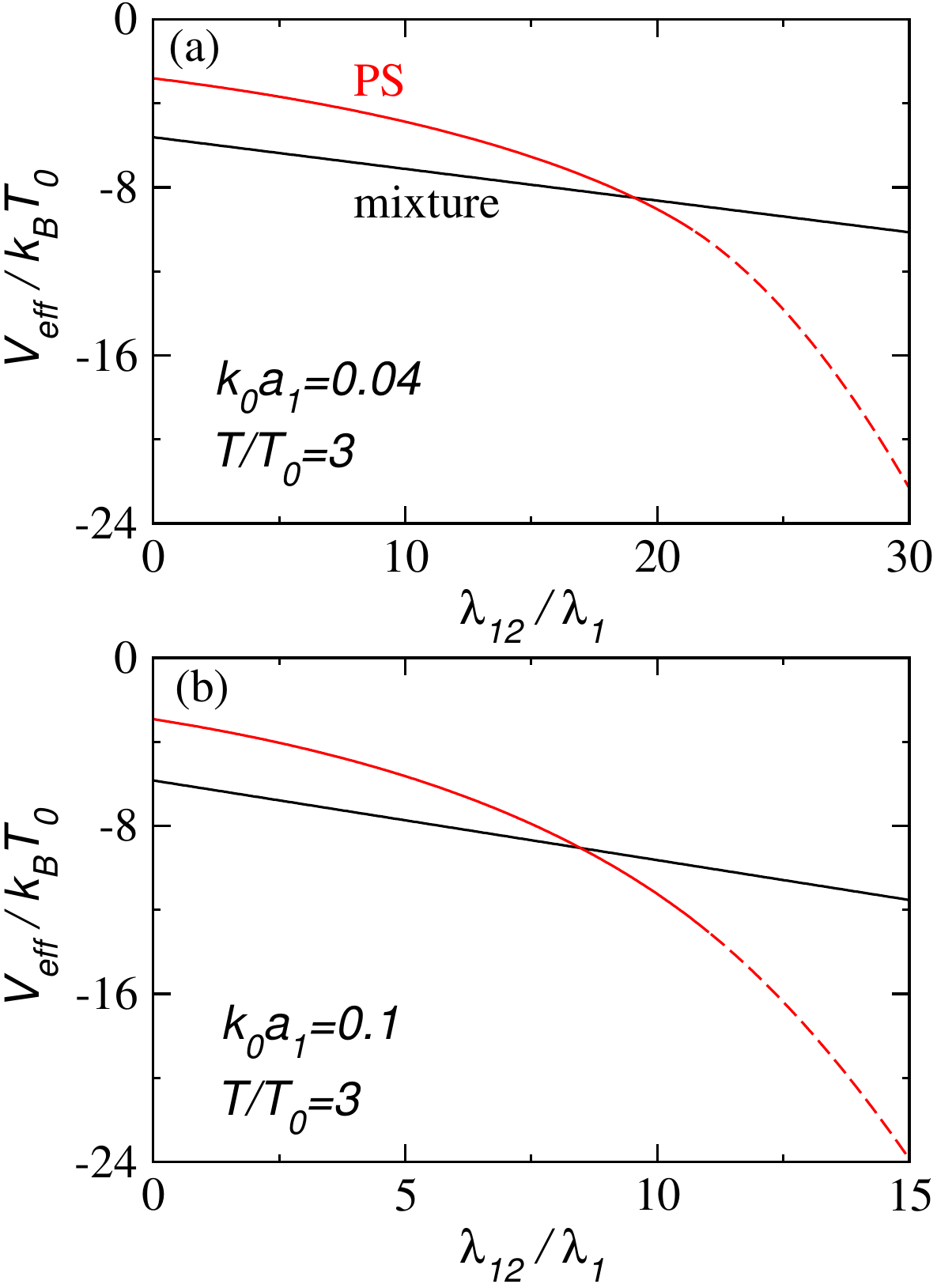}
  \caption{$V_{eff}$ as a function of $\lambda_{12}/\lambda_{1}$ at $T/T_0=3$ (evaluated at the mininum) for (a) $k_0 a_1=0.04$ and (b) $k_0 a_1=0.1$. Black (red) lines corresponding to the mixture state (phase-separated state denoted by PS). The dashed lines in the phase-separated state indicates that it is in the broken-symmetry phase. }
\label{fig:2BEC_Veff}
\end{figure} 
We now focus on the case where $\lambda_1=\lambda_2=4\pi\hbar^{2}a_1/m$.
Figure~\ref{fig:2BEC_Veff} shows $V_{eff}$ from the mixture state and phase-separated state at $T/T_0=3$ for two selected intra-species interaction strengths $k_0 a_1=0.04$ and $0.1$. For small $\lambda_{12}/\lambda_{1}$ the mixture phase is more stable due to its lower $V_{eff}$. As $\lambda_{12}/\lambda_1$ reaches a critical value, the two curves of $V_{eff}$ intersect and above the critical point the phase-separated state is more energetically  stable. In the grand-canonical ensemble implemented here, the two states are compared at \textit{the same chemical potentials}. Therefore the densities may not be the same in the two states. Note that when $\lambda_{12}/\lambda_1$ gets larger, the density in the phase-separated state increases in order to match the chemical potentials in the mixture state. Since there is no shift in $T_c$ from the leading-order single-auxiliary-field theory when compared to a noninteracting Bose gas, the critical temperature $T^{ps}_c=2\pi\hbar^{2}(\rho^{ps})^{2/3}/[\zeta^{2/3}(3/2)k_B m]$ of the phase-separated state  increases accordingly. Eventually the phase-separated state may enter the broken-symmetry phase if $T$ is not too high and we show this effect as the dashed lines in Fig.~\ref{fig:2BEC_Veff}.

\begin{figure}
  \includegraphics[width=3.in,clip] {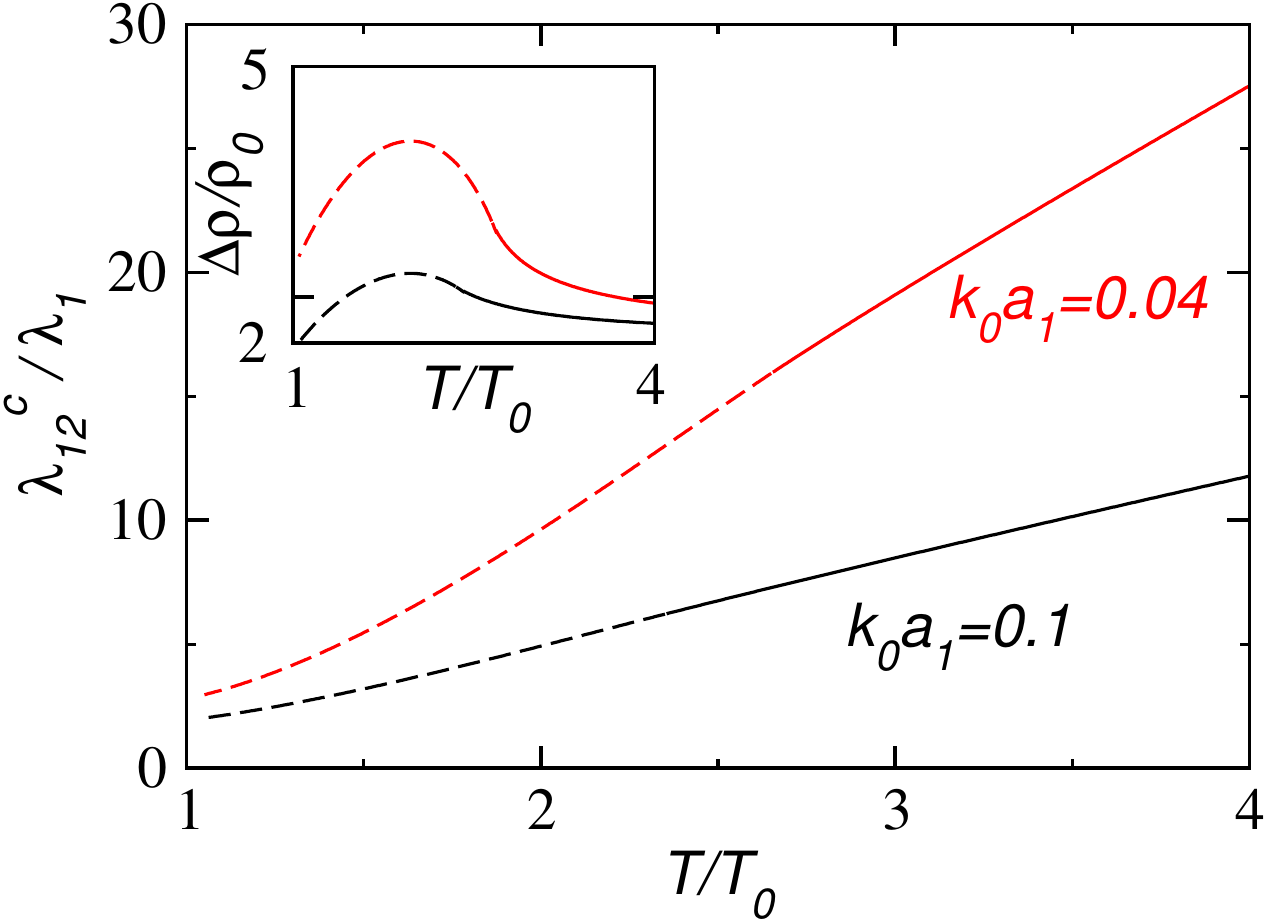}
  \caption{Phase diagram of the normal phase of a two-component Bose gas. The lines show the critical value of $\lambda_{12}/\lambda_1$ where a phase transition occurs for $k_0 a_1=0.1$ (black) and $0.04$ (red). Below the critical line the system is a mixture of normal gases and above the line a phase-separated state emerges. The dashed lines indicate that the phase-separated state is in a symmetry-broken phase. Inset: The normalized density difference $(\rho_{ps}-\rho)/\rho_0$ at the critical line.}
\label{fig:normal_phase}
\end{figure} 
By locating the critical value of $\lambda_{12}/\lambda_{1}$ where the two curves of $V_{eff}$ intersect at fixed $T$, we found the phase diagram shown in Figure~\ref{fig:normal_phase}. Each curve corresponds to the critical line separating the mixture state and the phase-separated state. One can see that the mixture state prefers lower $\lambda_{12}/\lambda_{1}$ and higher $T$ while the phase-separated state prefers the opposite. The dashed lines in Fig.~\ref{fig:normal_phase} indicates that the phase-separated state enters the broken-symmetry phase. In that regime the leading order in large-$N$ approximation is only a qualitatively accurate result. In the region near and below  $T_c$  one can use the LOAF theory \cite{LOAFPRL,LOAFlong} to improve on the result presented here since that approximation exactly reproduces Bogoliubov's results at weak coupling as well as correctly includes the anomalous density and predicts a shift in $T_c$ from the free-gas result. However, using LOAF  will not change the answer in the region when both states are in the normal phase and would unnecessarily complicate the simplicity of the calculation presented here. One could also include the next to leading order $1/N$ terms to be able to access the regime around $T_c$ where the anomalous density  correlations become important.  A structural transition from a homogeneous mixture state into a phase-separated state in the normal phase has also been studied in two-component Fermi gases with population imbalance \cite{ChienPRL97}. The underlying mechanisms are different: For fermions the system is maximizing the pairing energy while for bosons the system is minimizing the repulsive interactions.

Importantly, only the global translational symmetry is broken in the phase transition from the mixture phase to the phase-separated phase when  both species are in the normal phase. In the phase-separated phase, there is an interface separating the two components and each component respects the local translational symmetry away from the interface. The different densities of the two states across this mixture to phase-separation transition remind us of the liquid-vapor transition, where no symmetry is broken and the density difference serves as the "order parameter" distinguishing the two phases. We thus study the normalized density difference $(\rho^{ps}_{j}-\rho_{j})/\rho_0$ at the critical line in the inset of Fig.~\ref{fig:normal_phase}. If the particle number is conserved when one compares the two states, the normalized density difference should be $1$ because the density in the phase-separated state should be twice as large as that of the mixture state. However, since we are comparing the two states \textit{at the same chemical potential}, one can see from the inset of Fig.~\ref{fig:normal_phase} that the conservation of the particle number is not respected. 

To draw the phase diagram with particle number conservation, one has to work in the canonical ensemble with fixed particle numbers and find the corresponding free energy. The physics should be the same if the results are compared correctly. For an isolated atomic cloud, our instability analysis may apply to a small region with the rest of the cloud treated as a reservoir. The phase separation could start growing if the mixture state is unstable in that focused region and the instability may propagate to the whole cloud.

One has seen that the instability of a mixture of two-component Bose gases can be analyzed using the mean-field approximation derived from the  leading order in the  large-$N$ expansion, which involves the introduction of an auxiliary field related to the normal density. The detailed structures which develop when the system evolves into a phase-separated state, however, require numerical simulations of 
the equations resulting from the effective action and are beyond the scope of the present paper.  The width of the interface separating the two species may be estimated using a variational method related to the one used in the estimation of the width of the interface separating two BEC phases in the ground state discussed in Ref.~\cite{EddyPRL98}.

\section{Conclusion}\label{sec:conclusion}
We have shown that the leading order in our large-$N$ approximation, which utilizes a single auxiliary field related to the normal density, leads to a mean-field theory usable at all couplings and temperatures that is a valuable tool for investigating the physics of interacting Bose gases. For a single-component Bose gas we show that, by constructing the  propagators in the broken symmetry vacuum, a Bogoliubov-like dispersion indeed emerges. For a two-component Bose gases, this approximation predicts a normal-phase structural phase transition between a mixture state and a phase-separated state. One possible application of two-component Bose gases is to simulate cosmological dynamics \cite{Fischer04}. Our theory may help extend this application beyond the low-temperature regime.

The authors acknowledge the support of the U. S. DOE through the LANL/LDRD Program. C. C. C. and F. C. thank the hospitality of Santa Fe Institute.

\bibliographystyle{apsrev4-1}
%\bibliography{reference}
%merlin.mbs 2010-03-15 4.21a (PWD, AO, DPC)

\end{document}